\documentclass[conference]{IEEEtran}

\usepackage{cite}
\usepackage{rotating}
\usepackage{bm}
\usepackage{enumerate}
\usepackage[cmex10]{amsmath}
\usepackage{multirow}%
\usepackage{array}
\usepackage{mdwmath}
\usepackage{amssymb}
\usepackage{amsthm}
\usepackage[tight,footnotesize]{subfigure}
\usepackage{amsmath,amsthm}
\usepackage{threeparttable}
\usepackage{booktabs}
\usepackage{textcomp,booktabs}
\usepackage[usenames,dvipsnames]{color}
\usepackage{colortbl}
\usepackage[table]{xcolor}
\usepackage{threeparttable}

\usepackage{url}
\usepackage[printonlyused]{acronym}

\begin{document}

\title{Modeling and Analysis of MPTCP Proxy-based LTE-WLAN Path Aggregation}

\author{\IEEEauthorblockN{Bolin Chen\IEEEauthorrefmark{1},
Zheng Chen\IEEEauthorrefmark{2}, Nikolaos Pappas\IEEEauthorrefmark{3}, Di Yuan\IEEEauthorrefmark{3} and Jie Zhang\IEEEauthorrefmark{1}}
\IEEEauthorblockA{\IEEEauthorrefmark{1}Department of
Electronic and Electrical Engineering,
The University of Sheffield, Sheffield, UK\\
\IEEEauthorrefmark{2}Department of Electrical Engineering,
Link\"{o}ping University, Link\"{o}ping, Sweden \\
\IEEEauthorrefmark{3}Department of Science and Technology,
Link\"{o}ping University, Norrk\"{o}ping, Sweden \\
Email: \IEEEauthorrefmark{1}bchen16@sheffield.ac.uk,
\IEEEauthorrefmark{2}zheng.chen@liu.se}}
\maketitle

\maketitle

\begin{abstract}
Long Term Evolution (LTE)-Wireless Local Area Network (WLAN) Path Aggregation (LWPA) based on Multi-path Transmission Control Protocol (MPTCP) has been under standardization procedure as a promising and cost-efficient solution to boost Downlink (DL) data rate and handle the rapidly increasing data traffic.
This paper aims at providing tractable analysis for the DL performance evaluation of large-scale LWPA networks with the help of tools from stochastic geometry. We consider a simple yet practical model to determine under which conditions a native WLAN Access Point (AP) will work under LWPA mode to help increasing the received data rate. Using stochastic spatial models for the distribution of WLAN APs and LTE Base Stations (BSs), we analyze the density of active LWPA-mode WiFi APs in the considered network model, which further leads to closed-form expressions on the DL data rate and area spectral efficiency (ASE) improvement. Our numerical results illustrate the impact of different network parameters on the performance of LWPA networks, which can be useful for further performance optimization.
\end{abstract}

\IEEEpeerreviewmaketitle

\section{Introduction}
There is an increasing demand for high data rate in wireless communications systems due to the fact that mobile traffic is expected to become 8.2 times larger in 2020 than what it was in 2015. Since the new licensed spectrum bands are rare and expensive, an interesting proposition is to enable a better use of different types of spectrum traffic offload, including the unlicensed bands. It is estimated that up to thirty percent of broadband access in cellular networks can be offloaded to unlicensed bands which are primarily used by WiFi networks \cite{In1}.

There have been attempts to develop Long Term Evolution (LTE)-Wireless Local Area Network (WLAN) Path Aggregation (LWPA) based on Multi-path Transmission Control Protocol (MPTCP) \cite{BPB11,kaspar2012multipath}. The LWAP design allows the aggregation of LTE and WiFi, and the increase of download speeds can be acquired simply by adding an MPTCP gateway on mobile device, without causing changes in legacy Internet infrastructure or applications. Unlike LTE-WLAN carrier aggregation solutions like the LTE-Licensed Assisted Access (LAA), LWAP eliminates the controversial fairness issues between LTE and WiFi and it does not require the deployment of new access networks\cite{me1}. LWPA can be potentially incorporated with Quality-of-Service (QoS)-related applications, such as Scalable Video Coding (SVC). For instance, the basic layer data can be delivered through a reliable LTE connection and the enhancement layer data can be transmitted through a WiFi Access Point (AP) to increase the quality of video streaming.

A mathematical approach would be helpful for a fundamental understanding of the performance of LWPA. Recently, stochastic geometry has become a powerful tool for modeling cellular and WiFi systems with large random topologies. In \cite{PPP1}, the coverage probability and the average Shannon rate were derived for macro cellular networks with Base Stations (BSs) distributed according to a \textit{Poisson Point Process} (PPP). This work has been extended to heterogeneous cellular networks (HetNets) \cite{Hetnet1, Hetnet2, Hetnet3}. Stochastic geometry can also model Carrier Sense Multiple Access with Collision Avoidance (CSMA/CA)-based WiFi networks. When using CSMA/CA as the MAC protocol, due to the carrier sensing range, there is a minimum guaranteed distance between concurrent transmitters operating in the same frequency band. \textit{Mat\'{e}rn Hard-core Process} (MHCP) has become a commonly used model for the distribution of simultaneously transmitting nodes where two points cannot coexist with a separating distance shorter than a predefined parameter \cite{mean_interference, bacelli_stogeo, Yuan}. Thus, MHCP can capture the repulsion between two coexisting nodes. Another commonly used model for networks with inter-tier repulsion is the \textit{Poisson Hole Process} (PHP), which can capture the inter-tier correlation when a secondary node cannot be within a predefined distance to a primary node \cite{PHP1, PHP2, ISWCS2014, d2d_journal}. In a Device-to-Device (D2D) underlaid cellular network where the activation conditions of D2D nodes depend on the locations of neighbors in both cellular and D2D tiers \cite{ICASSP2014}, the effect of two-tier dependence on the network throughput has been characterized in \cite{two_tier_chen}.

In this work, a stochastic geometry framework is proposed to evaluate the performance of LWPA in large-scale networks. We take into account the possibility to have closed-access WiFi APs that are not available to be aggregated and whose performance are not allowed to be affected. The activation conditions of a LWPA-mode WiFi are determined by the user density, the CSMA distance, and their relevant locations to nearby closed-access WiFi APs. Using existing results from stochastic geometry analysis, we propose three approximations for the density of active LWPA WiFi as a function of different network parameters. We also characterize the performance improvements of the considered network in terms of the aggregate data rate and the area spectral efficiency (ASE) of a WiFi band. These improvements validate the advantages of LWPA as a promising method to improve the spectrum usage in unlicensed bands.

\section{System Model}
\label{sec:system}
We consider a two-Radio Access Technology (RAT) network consisting of cellular BSs and WLAN APs. The LTE BSs are scattered on the two-dimensional Euclidian plane $\mathbb{R}^2$ according to a homogeneous PPP, denoted by $\Phi_L$ with intensity $\lambda_L$ \cite{haenggi2012stochastic}. The coverage region of an LTE BS can be modeled via a Voronoi tessellation. Similarly, the WLAN APs are modeled to follow a homogeneous PPP $\Phi_{W}$ with intensity $\lambda_{W}$. We divide the WLAN APs into two tiers/groups, depending on whether or not they are accessible for the network operators to improve the QoS of cellular networks. The open-access WiFi APs are available for path aggregation with the LTE BSs to serve cellular users simultaneously through parallel data flows. The closed-access WiFi are occupied by private individuals, hence, they are accessible to only WiFi users with the access rights. We assume that each WiFi AP has a probability $p$ to be occupied by closed-access WiFi users, and probability $1-p$ to be open-access.\footnote{The probability $p$ denotes the percentage of closed-access WiFi among all the WiFi APs.} As a result, the WiFi locations can also be modeled as two independent PPPs $\Phi_{W1}$ and $\Phi_{W2}$ for the open-access and closed-access WiFi tiers, respectively, with corresponding densities $\lambda_{W1}=(1-p)\lambda_{W}$ and $\lambda_{W2}=p\lambda_{W}$. The LTE users are also assumed to be distributed according to another homogeneous PPP $\Phi_u$ with density $\xi_u$. We suppose that each cellular user is associated with the closest LTE BS. The distribution of users connected to closed-access WiFi will not affect the performance of LWPA, thus it is not specified in our system model.

For the LTE cellular network, each LTE BS utilizes the total available spectrum of $B_c$-Hz universal frequency reuse in the Downlink (DL), which is partitioned into a number of radio Resource Blocks (RBs). Each RB occupies a bandwidth of $b_c$ Hz, and
the RBs are allocated equally among all users. Assuming having $N$ users inside a LTE cell, then each user is assigned $B_c/b_cN$ RBs. The WiFi APs utilize the unlicensed spectrum with a total bandwidth of $B_w$ Hz.

\begin{figure}[!t]
\centering
\includegraphics[width=\linewidth]{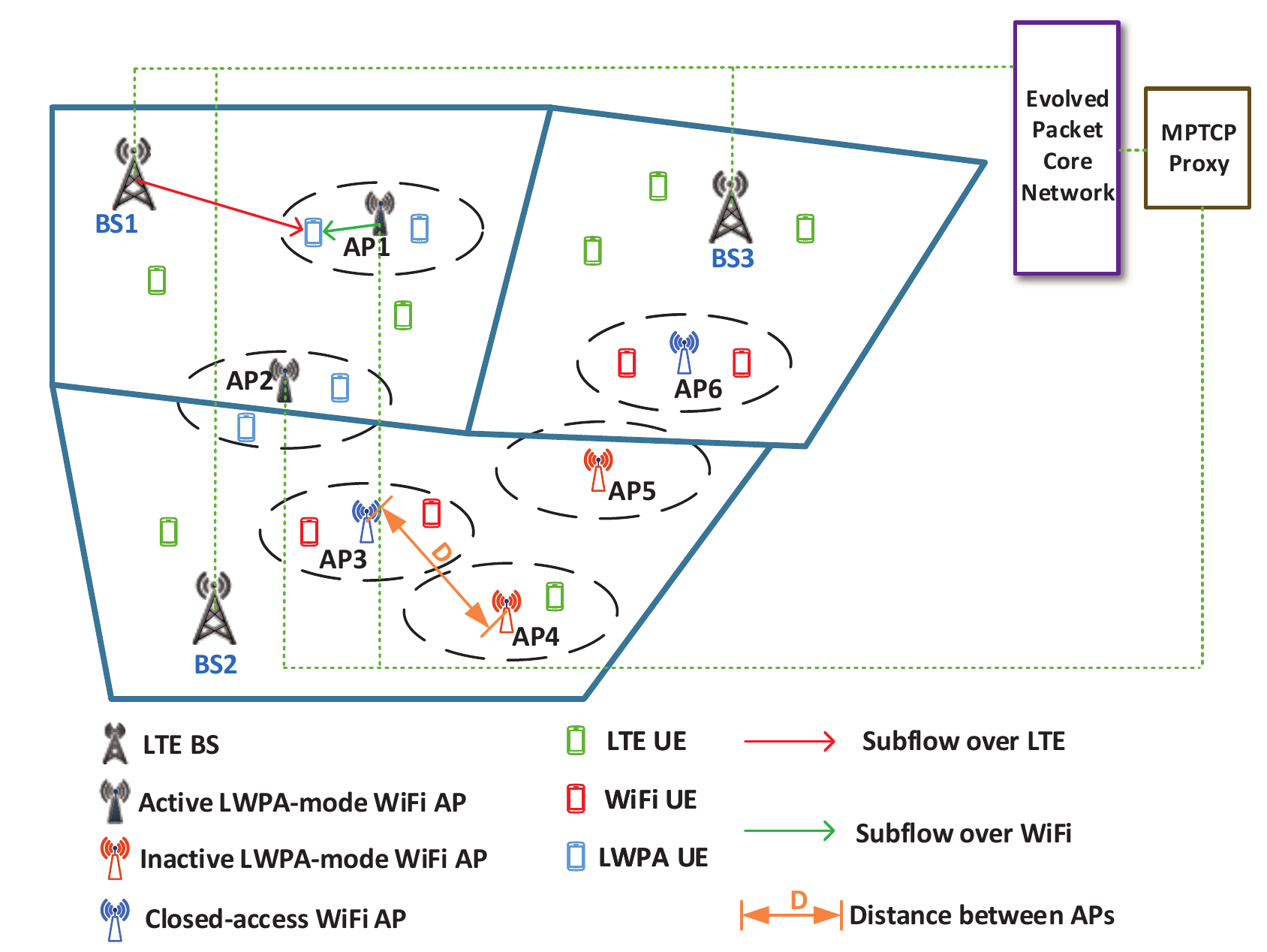}
\caption{Two-RAT HetNet consisting of LTE BSs, active LWPA-mode WiFi APs, inactive LWPA-mode WiFi APs in the cellular RAT, and closed-access WiFi APs in the WLAN RAT. The backhaul is shown by the dotted lines. AP1 and AP2 are active LWPA-mode WiFi APs since they both meet the conditions. AP3 and AP6 are not available for LWPA mode because they are closed-access WiFi. Although AP4 and AP5 are open-access WiFi, they are inactive LWPA-mode WiFi APs because there is no cellular user closer to AP5 than distance $R$, while the distance between AP4 and the active WiFi AP3 is shorter than $\delta$.}
\label{System}
\end{figure}

In this work, we consider MHCP of type II to model the distribution of WiFi APs using CSMA-CA MAC protocol \cite{bacelli_stogeo}. The key point is that two active WiFi APs operating in the same frequency band cannot be closer to each other than a threshold distance $\delta$, which can be seen as setting up guard zones with a radius $\delta$ around the active transmitters.
The distribution of active closed-access WiFi APs without the presence of LWPA-mode WiFi is denoted by $\widetilde{\Phi}_{W2}$. In order to ensure the QoS of closed-access WiFi users, the activation of LWA-mode WiFi APs must not create any backoff/contention to any active closed-access WiFi.
According to this setup, a potential LWPA-mode WiFi AP can be active only when there is no closed-access WiFi AP or other active LWPA-mode WiFi AP within distance $\delta$.
The distribution of active LWPA-mode WiFi APs is a thinned version of the initial homogeneous PPP $\Phi_{W1}$, where the thinning procedure involves conditions imposed by the cellular user density, the locations of closed-access WiFi and other LWA-mode WiFi APs. Fig. \ref{System} presents a two-RAT HetNet consisting of LTE BSs in the cellular RAT and active LWPA-mode WiFi APs, inactive LWPA-mode WiFi APs, as well as closed-access WiFi APs in the WLAN RAT.

Summarizing, an active LWPA-mode WiFi AP must satisfy the four conditions below:
\begin{enumerate}
\item It must be open-access;
\item At least one LTE user is inside its service range, i.e., within distance $R$;
\item Any closed-access WiFi AP must be at least outside distance $\delta$;
\item Two active LWPA-mode WiFi APs cannot be closer to each other than $\delta$.
\end{enumerate}
When all these conditions are satisfied, an open-access WiFi AP can work in LWPA mode and serve the cellular user within its coverage together with the LTE BS.

Using existing results on the density of MHCP and Poisson Hole Process (PHP), we can obtain the following approximations on the density of active LWPA-mode WiFi APs in our considered model.

\begin{enumerate}[(a)]
\item From the first and the second conditions, the probability of a point in $\Phi_{W1}$ being retained is the probability that there is at least one point from $\Phi_u$ within distance $R$. The retaining probability is \cite{haenggi2012stochastic}
     \begin{equation}
    P_1={1-\exp(-\xi_u\pi R^2)}.
    \end{equation}
     Considering the CSMA/CA operation of the WLAN RAT, the WiFi APs in the same tier are assumed to be installed with respect to a minimum distance $\delta$ between each other, thus the resulted distribution of closed-access WiFi APs forms an MHCP $\widetilde\Phi_{W2}$ with intensity $\widetilde\lambda_{W2}=(1-\exp(-p\lambda_W\pi\delta ^2))/{\pi\delta ^2}$ \cite{mean_interference}. Based on the first and the third conditions, the probability of a point in $\Phi_{W1}$ being retained is the probability that there is no point from $\widetilde \Phi_{W2}$ within distance $\delta$. The resulted intensity of retained points can be approximated by the density of a Poisson Hole Process (PHP). The corresponding retaining probability is
     \begin{equation}
     \begin{aligned}
     P_2= & \exp(-\widetilde \lambda_{W2}\pi \delta ^2)\\
     =&\exp(\exp(-p\lambda_W\pi\delta ^2)-1).
     \end{aligned}
    \end{equation}
   Note that this approximation is accurate only when $\widetilde \lambda_{W2}<P_1\lambda_{W1}$ \cite{PPP3}.
      Thus, the distribution of open-access WiFi APs from $\Phi_{W1}$ meeting the second and third conditions forms $\widetilde\Phi_{W1}$ with intensity $\widetilde\lambda_{W1}=P\lambda_{W1}$, where the retaining probability is $P=P_1P_2$. Because of the fourth condition the resulted distribution of active LWPA-mode WiFi APs can be further regarded as an MHCP $\Phi_{A}^1$ with intensity \cite{mean_interference}

\begin{equation}\label{Lemma1.1}
\lambda_{A}^1\approx\frac{1-\exp(-\widetilde\lambda_{W1}\pi\delta ^2)}{\pi \delta ^2}.
\end{equation}

\item Due to the fourth condition, we have that the resulted distribution of open-access WiFi APs based on $\Phi_{W1}$ forms an MHCP $\widehat\Phi_{W1}$ with intensity $\widehat\lambda_{W1}=\frac{1-\exp(-\lambda_{W1}\pi \delta ^2)}{\pi \delta ^2}$. Thus, the active LWPA-mode WiFi APs refer to those from $\widehat\Phi_{W1}$ meeting the second and third conditions. The effective density of active LWPA-mode WiFi APs modeled as $\Phi_{A}^2$ is
\begin{equation}\label{Lemma1.2}
\lambda_{A}^2\approx P_1 P_2\widehat\lambda_{W1},
\end{equation}
when $\widetilde \lambda_{W2}<P_1\widehat\lambda_{W1}$ holds. $\Phi_{A2}$ is obtained by thinning the MHCP $\widehat\Phi_{W1}$ with the retaining probability $P=P_1 P_2$.

\item We first assume that the distribution of WiFi APs meeting the fourth condition forms an MHCP $\widetilde\Phi_{W}$ with intensity $\widetilde\lambda_{W}=\frac{1-\exp(-\lambda_{W}\pi \delta ^2)}{\pi \delta ^2}$. Then, the intensity of open-access WiFi APs from $\widetilde\Phi_{W}$ can be written as $(1-p)\widetilde\lambda_{W}$. The active LWPA-mode WiFi APs refer to the open-access WiFi APs from $\widetilde\lambda_{W}$ meeting the second condition, and they are modeled as $\Phi_{A}^3$ with intensity
\begin{equation}\label{Lemma1.3}
\lambda_{A}^3\approx P_1(1-p)\widetilde\lambda_{W},
\end{equation}
which is obtained by thinning the MHCP $\widetilde\Phi_{W}$ with the retaining probability $(1-p)P_1$.
\end{enumerate}

\begin{figure}[!t]
\centering
\includegraphics[width=\columnwidth]{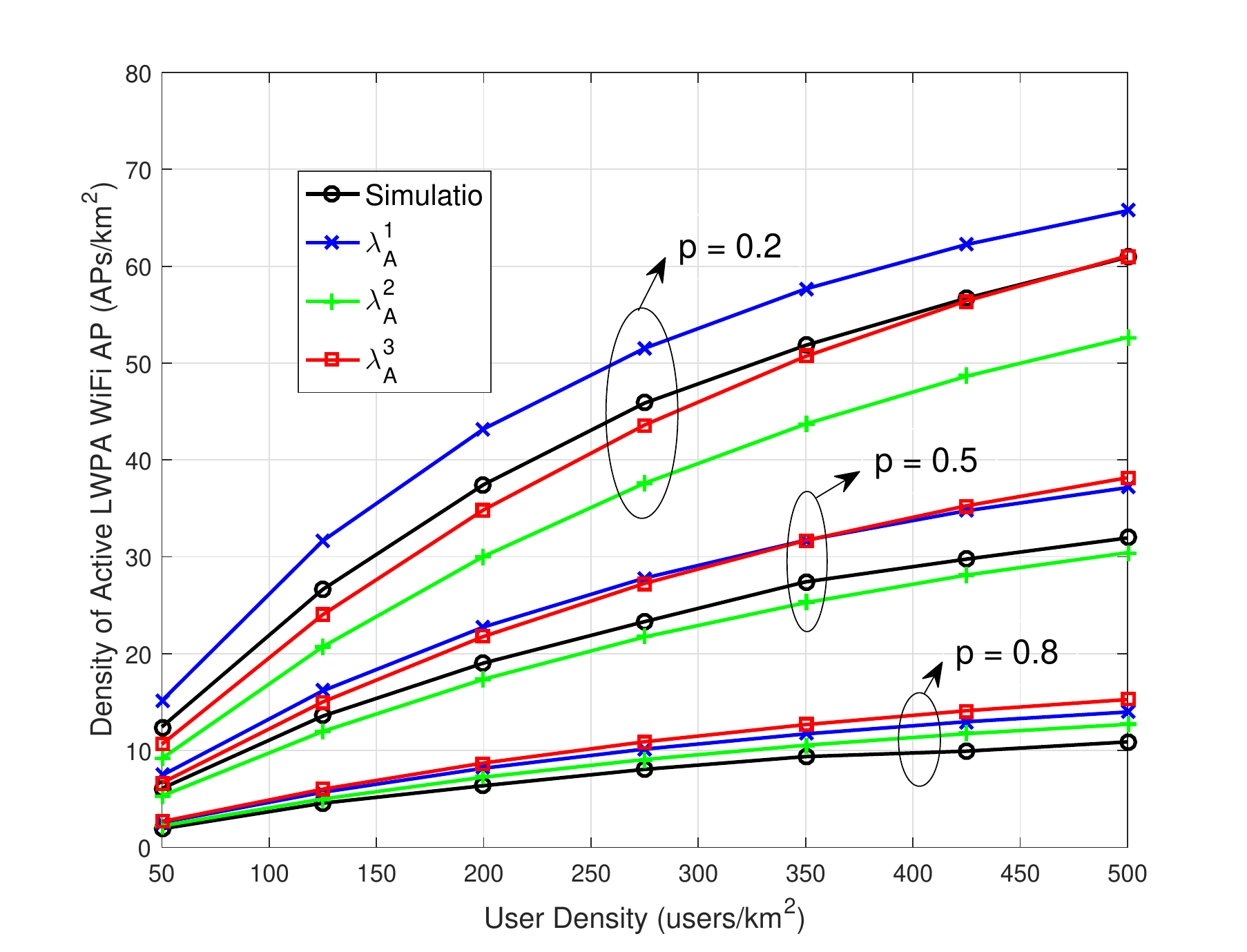}
\caption{Density of active LWPA WiFi APs vs. user density with $p=0.2, 0.5, 0.8$ and $\delta=50$ $\mathrm{m}$.}\label{LWAdensity}
\end{figure}

Fig. \ref{LWAdensity} shows the density of active LWPA-mode WiFi APs versus the user density in cases when $p$ equals to $0.2$, $0.5$ and $0.8$, respectively. The approximated density is calculated with $\lambda_L=100\mathrm{/km^2}$, $\lambda_{W}=200\mathrm{/km^2}$ and $R=30$ $\mathrm{m}$. The results shown in blue lines are generated from (\ref{Lemma1.1}). The red lines correspond to the approximation results obtained via (\ref{Lemma1.2}). The green lines refer to the results generated via (\ref{Lemma1.3}). 
We can observe from the results that the three approximations are all relevant, and the accuracy of these approximations depends heavily on the network parameters, such as $\delta$ and $p$.
The gaps between the approximations and simulation results also come from the fact that the existing density functions of MHCP and PHP also contain certain approximation errors.
For simplicity, in the remainder of this paper, we consider $\lambda_{A}=\lambda_{A}^3$ to represent the density of active LWPA-mode WiFi APs.

\section{Performance Analysis}
In this section, we evaluate the performance gain of LWPA, in terms of the improved data rate and the area spectral efficiency.
\subsection{Success Probability Analysis}
In this work, we assume that in each time slot, at most one user will be served by an LWPA-mode WiFi. Due to the randomness of point processes, for each active LWPA-mode WiFi, its served cellular user can be considered as randomly and uniformly distributed within its service range.
Without loss of generality, we consider a typical LWPA user centered at the origin with its associated LWPA-mode WiFi AP $W_o$ at a random distance $l$ away, the SINR of its received signal is given by
\begin{equation}
\mathrm{SINR}_W=\frac{P_Wh_Wl^{-\alpha}}{\sum _{j\in\Phi_{A} \cup \widetilde\Phi_{W2} \setminus\{W_o\}}P_Wh_jd_{j}^{-\alpha }+\sigma ^2},
\end{equation}
where $P_W$ denotes the transmitting power of WiFi AP. $h_W$ and $h_j$ refer to small-scale fading from the typical LWPA-mode WiFi AP, and the $j$-th interfering WiFi AP to the typical LWPA user, respectively. $\sigma ^2$ represents the noise power. We assume that all users experience Rayleigh fading, i.e. $h_W, h_j\sim \exp(1)$. $d_{j}$ denotes the distance from the $j$-th interfering WiFi AP to the typical LWPA user. We consider a standard distance-dependent pathloss attenuation, i.e. $r^{-\alpha}$, where $\alpha>2$ is the pathloss exponent.

Similarly, for a typical LTE user located at the origin with its associated LTE BS $L_o$ at a random distance $r$ away, the SINR of its received signal is given by
\begin{equation}
\mathrm{SINR}_L=\frac{P_Lg_Lr^{-\alpha}}{\sum _{i\in\Phi_{L}\setminus\{M_o\}}P_Lg_il_{i}^{-\alpha}+\sigma ^2},
\end{equation}
where $P_L$ denotes the transmit power of LTE BS. $g_L$ and $g_i$ refer to small-scale power fading from the typical LTE BS, and the $i$-th interfering LTE BS to the typical LTE user, respectively. $g_L$ and $g_i$ follow the exponential distribution with unit mean (Rayleigh fading). $l_{i}$ denotes the distance from the $i$-th interfering WiFi AP to the typical LTE user.

\begin{figure}[!t]
	\centering
	\includegraphics[width=2.5in]{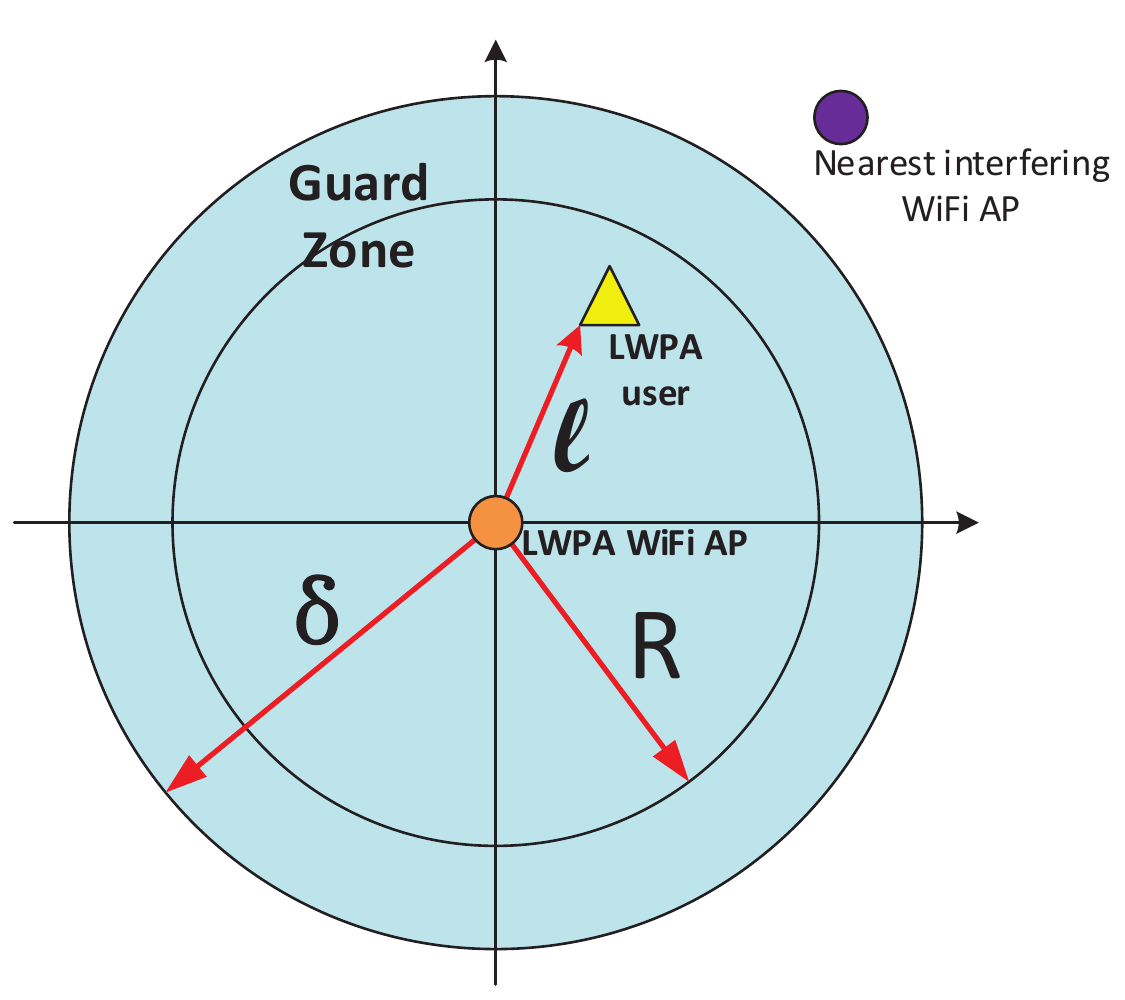}
	\caption{Guard Zone with radius $\delta$ of a typical active LWPA-mode WiFi AP centered at the origin with its users uniformly distributed in a disk $B(0,R)$. According to the system model, no active open-access or closed-access WiFi AP can lie in the circle of radius $\delta$ centered at the origin, thus the distance from the nearest WiFi AP to the typical LWPA user is at least $\delta-l$.}\label{Guardzone}
\end{figure}

As discussed in Section \ref{sec:system}, the interference received by the typical LWPA user is caused the transmitting nodes distributed outside the guard zone centered at its associated transmitter. When $\delta \ge l$, there is a minimum distance between a typical LWPA user to its nearest interfering WiFi AP, as shown in Fig. \ref{Guardzone}. The probability density function (PDF) of the link distance $l$ is $f(l)=\frac{2l}{R^2}$ for $0\leq l\leq R$, as a result of the randomly and uniformly distributed user within the service range of an LWPA-mode  WiFi AP.
For a given SINR threshold $\theta$, the success probability of a typical active LWPA-mode WiFi link in our considered network is given by
\begin{equation}\label{Wifisuc}
\begin{aligned}
P_{suc}^{WiFi}(\theta, \lambda)=&\mathbb{P}(\mathrm{SINR}_W>\theta)\\
                     =&\mathbb{E}_l\Bigg[\mathcal{L}_{I_W|\delta-l}(\theta l^{\alpha}) \cdot \exp\Bigg(-\frac{\theta \sigma^2 l^{\alpha}}{P_W}\Bigg)\Bigg]\\
                     \approx &\int_{0}^{R}\frac{2l}{R^2} \cdot \mathcal{L}_{I_W|\delta-l}(\theta l^{\alpha}) \cdot \exp\Bigg(-\frac{\theta \sigma^2 l^{\alpha}}{P_W}\Bigg)\mathrm{d}l,
\end{aligned}
\end{equation}
where $I_W=\sum \limits_{j\in\Phi_{A} \cup \widetilde\Phi_{W2} \setminus\{W_o\}}h_jd_{j}^{-\alpha}$ denotes the interference from other active WiFi APs with normalized transmit power, $\mathcal{L}_{I_W|x}(s)$ refers to the Laplace transform of the interference coming from other WiFi APs located out of a circle area $B(0,x)$ and is given by \cite{two_tier_chen}
\begin{equation}
\begin{aligned}
\mathcal{L}_{I_W|x}(s)=&\mathbb{E}\Bigg[\exp\Bigg(-s\sum _{j\in\Phi_{A}\setminus B(0,x)}h_jd_j^{-\alpha}\Bigg)\Bigg]\\
                     = &\exp\Bigg(-\pi\lambda s^{\frac{2}{\alpha}} \int_{\frac{x^2}{s^{2/\alpha}}}^{\infty}\frac{1}{1+w^{\frac{\alpha}{2}}}\mathrm{d}w\Bigg),
\end{aligned}
\end{equation}
where $\lambda=\lambda_A+\widetilde\lambda_{W2}$.

As a result of the nearest LTE BS association, the PDF of a typical LTE link distance is $f(r)=2\pi\lambda_L r \cdot e^{-\pi\lambda_L r^2}$ for $0\leq r<\infty$\cite{PPP1}. Similarly, for a given SINR threshold $\theta$, the success probability of a typical LTE link in our considered network is given by
\begin{equation}
\label{eq:psuc_lte}
\begin{aligned}
P_{suc}^{LTE}(\theta)=&\mathbb{P}(\mathrm{SINR}_L>\theta)\\
                     =&\mathbb{E}_r\Bigg[\widetilde{\mathcal{L}}_{I_L|r}(\theta r^{\alpha}) \cdot \exp\Bigg(-\frac{\theta \sigma^2 r^{\alpha}}{P_L}\Bigg)\Bigg]\\
                     =&\int_{0}^{\infty}2\pi\lambda_L r \cdot e^{-\pi\lambda_L r^2} \cdot \widetilde{\mathcal{L}}_{I_L|r}(\theta r^{\alpha})\\
                      &\times \exp\Bigg(-\frac{\theta \sigma^2 r^{\alpha}}{P_L}\Bigg)\mathrm{d}r,
\end{aligned}
\end{equation}
where $I_L=\sum _{i\in\Phi_{L}\setminus\{M_o\}}g_il_{i}^{-\alpha}$ denotes the interference from other LTE BSs with normalized transmit power, $\widetilde{\mathcal{L}}_{I_L|x}(s)$ refers to the Laplace transform of interference coming from other LTE BSs located out of a circle area $B(0,x)$, given by \cite{two_tier_chen}
\begin{equation}
\begin{aligned}
\widetilde{\mathcal{L}}_{I_L|x}(s)=&\mathbb{E}\Bigg[\exp\Bigg(-s\sum _{i\in\Phi_{L}\setminus B(0,x)}g_il_i^{-\alpha}\Bigg)\Bigg]\\
                    =&\exp\Bigg(-\pi\lambda_{L} s^{\frac{2}{\alpha}} \int_{\frac{x^2}{s^{2/\alpha}}}^{\infty}\frac{1}{1+w^{\frac{\alpha}{2}}}\mathrm{d}w\Bigg).
\end{aligned}
\end{equation}

\subsection{Cellular Rate Improvement}
Without LTE-WLAN path aggregation, assuming i.i.d. Gaussian codebooks, the ergodic rate of a LTE link can be given by
\begin{equation}
\begin{aligned}
R_{LTE}=&\mathbb{E}[\log_2(1+\mathrm{SINR}_{L})]\\
       =&\int_{0}^{\infty}\frac{P_{suc}^{LTE}(\theta)}{1+\theta}\mathrm{d}\theta,
\end{aligned}
\end{equation}
which can be obtained with the help of the SIR distribution given in \eqref{eq:psuc_lte}.
Similarly, when the user is connected to a nearby LWPA-mode WiFi AP, the ergodic rate can be obtained by
\begin{equation}
R_{WiFi}=\int_{0}^{\infty}\frac{P_{suc}^{WiFi}(\theta, \lambda)}{1+\theta}\mathrm{d}\theta.
\end{equation}

As mentioned in Section~\ref{sec:system}, the RBs are equally assigned to the LTE users. In an LTE cell with total available bandwidth $B_c$, regardless of the number of users inside its coverage, the aggregate data rate averaging over all possible locations of users can be obtained as $B_c R_{LTE}$. With path aggregation, in addition to the data rate received from LTE BS, each active LWPA-mode WiFi AP can provide an average rate equal to $B_w R_{WiFi}$ to the LTE users inside its service range.
Therefore, in a random LTE cell, the percentage of cellular rate improvement can be expressed as
\begin{equation}\label{cellularrate}
P_{CI}=\frac{B_w \cdot R_{WiFi} \cdot N_W}{B_c \cdot R_{LTE}}.
\end{equation}
where $N_W=\frac{\lambda_A}{\lambda_L}$ is the average number of active LWPA-mode WiFi APs per LTE cell.

\subsection{Area Spectral Efficiency Improvement of the WiFi Band}
The ASE is an important metric to evaluate the spatial reuse of spectrum, measured by average data rate per Hz per unit area. For the considered network model, the ASE for WiFi spectrum can be expressed as
\begin{equation}
\begin{aligned}
\mathcal{T}
           =&(\lambda_A+\widetilde\lambda_{W2})\mathbb{E}[\log_2(1+\mathrm{SINR}_{W})]\\
           =&(\lambda_A+\widetilde\lambda_{W2})\int_{0}^{\infty}\frac{P_{suc}^{WiFi}(\theta,\lambda_A+\widetilde\lambda_{W2})}{1+\theta}\mathrm{d}\theta.
\end{aligned}
\end{equation}

Specifically, if $\lambda_A=0$, i.e. there is no active LWPA-mode WiFi in the network, the ASE for WiFi spectrum can be expressed as
\begin{equation}
\widetilde{\mathcal{T}}=\widetilde\lambda_{W2}\int_{0}^{\infty}\frac{P_{suc}^{WiFi}(\theta,\lambda_{W2})}{1+\theta}\mathrm{d}\theta.
\end{equation}

Thus, the ASE improvement of the WiFi band is
\begin{equation}\label{ase}
P_{SI}=\frac{\mathcal{T}}{\widetilde{\mathcal{T}}}.
\end{equation}
The improvement/ratio of ASE provides a quantitative measure on how much network throughput gain in WiFi band we can expect by allowing LWPA without affecting the activity of closed-access WiFi APs.

\section{Numerical Results}
In this section, we present the LWPA-mode WiFi link success probability, the cellular rate and the ASE improvement of the WiFi band through numerical evaluations for $p =\{0.2, 0.5, 0.8\}$.
The densities of the LTE BSs and WiFi APs are $\lambda_L=100\mathrm{/km^2}$ and $\lambda_W=200\mathrm{/km^2}$, respectively. The cellular and WLAN bandwidth are both set to be $10$ MHz. The transmit power of LTE BS and WiFi AP are set to be $22$ dBm and $18$ dBm, respectively, and the noise power on both licensed and unlicensed bands is $-95$ dBm. Rayleigh fading model is adopted for both cellular and WiFi links with $\mathbb{E}[h]=1$. The pathloss exponent is $\alpha=4$.

\begin{figure}[!t]
\centering
\includegraphics[width=0.9\columnwidth]{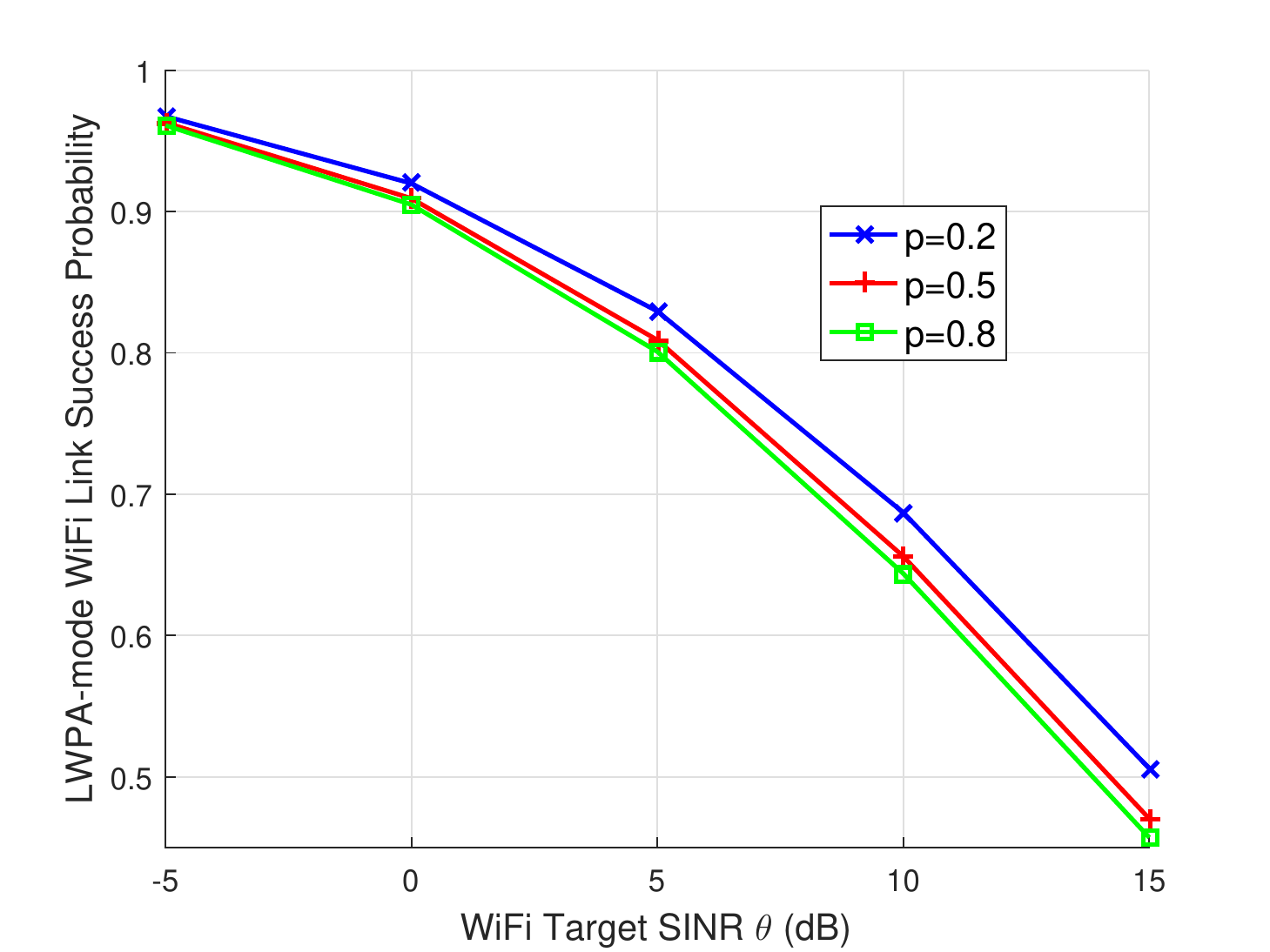}
\caption{LWPA-mode WiFi link success probability vs. $\theta$ with $p = \{0.2, 0.5,0.8\}$, $\delta=50$ m.}\label{LWAsuc2}
\end{figure}

\subsection{LWPA-mode WiFi Link Success Probability}
In Fig. \ref{LWAsuc2}, we show the approximated success probability of the LWPA-mode WiFi link as a function of the SINR threshold $\theta$. The results are generated from (\ref{Wifisuc}). We observe that the value of $p$ has little impact on the success probability of LWPA-mode WiFi link. With smaller closed-access WiFi percentage $p$, the success probability is slightly higher. This means that in the case with smaller $p$, the total density of active WiFi will be slightly smaller.

\begin{figure}[!t]
	\centering
	\includegraphics[width=0.9\columnwidth]{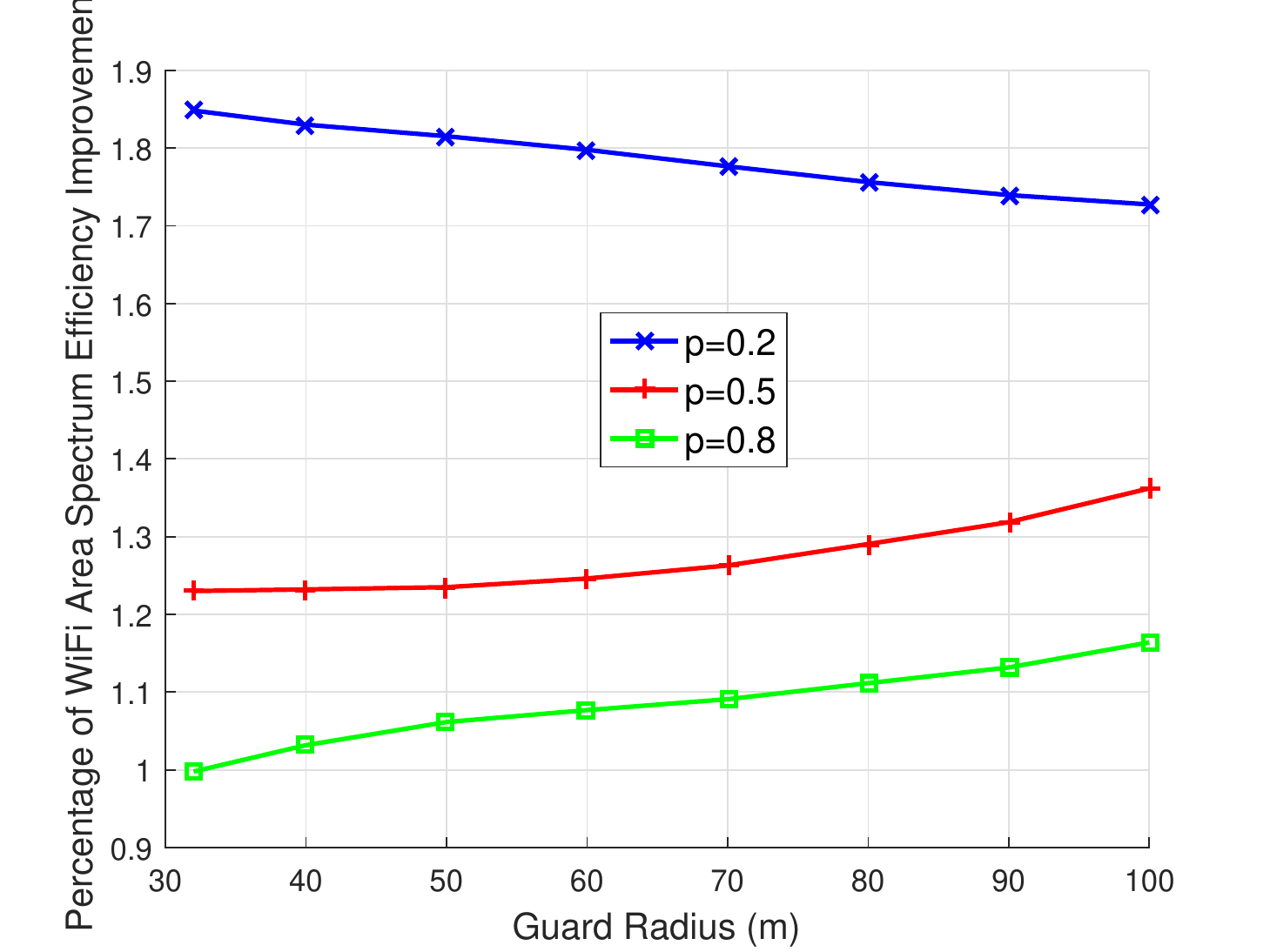}
	\caption{WiFi ASE improvement vs. guard zone radius with $p =\{0.2, 0.5, 0.8\}$.}\label{ASEradius}
\end{figure}

\begin{figure}[!t]
	\centering
	\includegraphics[width=0.9\columnwidth]{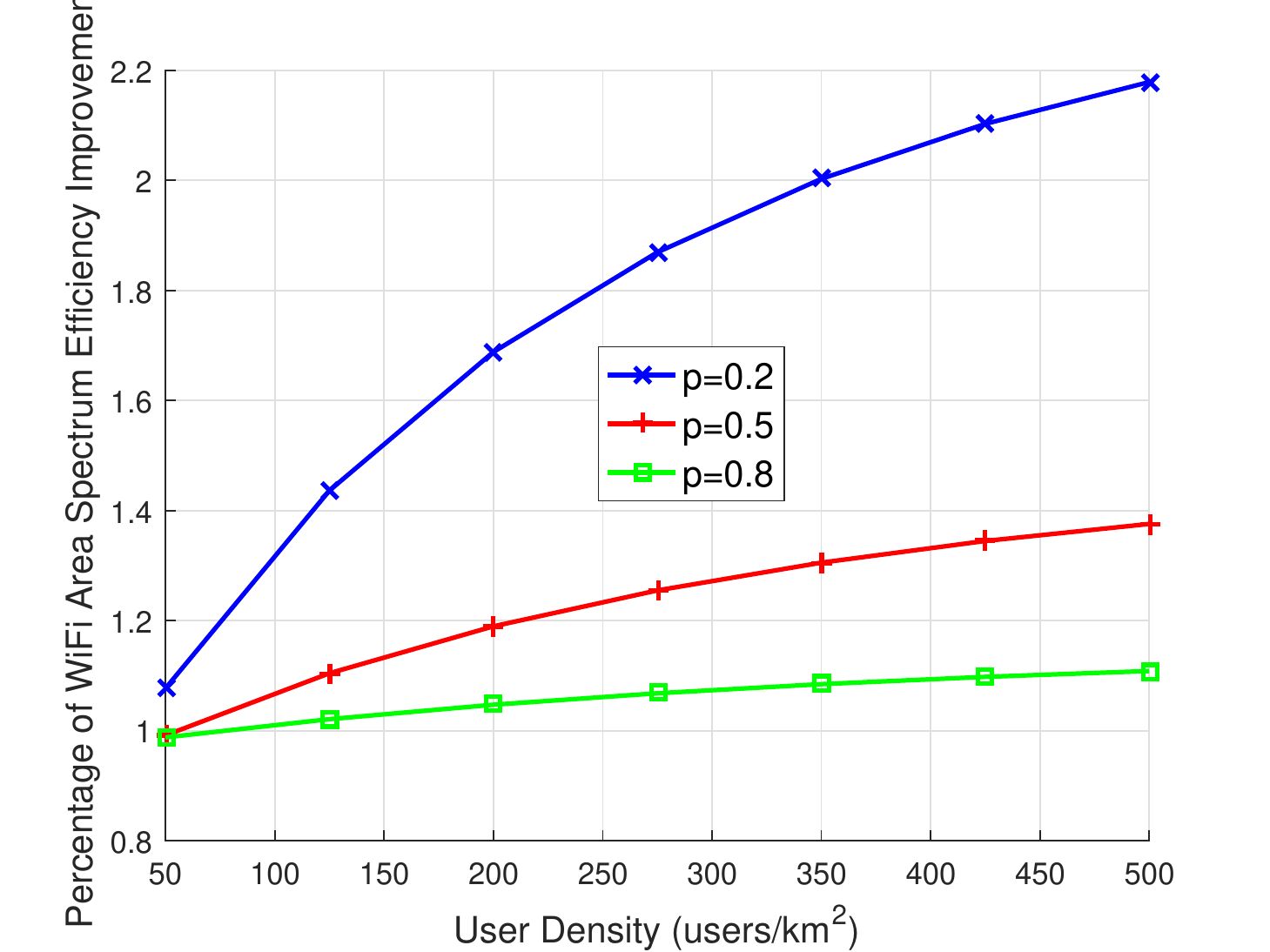}
	\caption{WiFi ASE improvement vs. user density with $p =\{0.2, 0.5, 0.8$\}, and here $\delta $= 50 m.}\label{ASEuser}
\end{figure}

\subsection{WiFi ASE Improvement}
Based on the approximated LWPA-mode WiFi link success probability, we present the the percentage of WiFi ASE improvement versus the guard zone radius as well as the user density in Fig. \ref{ASEradius} and Fig. \ref{ASEuser}, respectively. The results are generated from (\ref{ase}). As expected, with smaller closed-access WiFi percentage $p$, the WiFi ASE improvement is higher.
Another interesting observation from Fig. \ref{ASEradius} is that when $p= 0.5$ and $0.8$, the percentage of improved ASE increases along with the guard radius, while when $p= 0.2$, a different trend can be observed. This is due to the fact that increasing the guard radius will decrease the density of not only active LWPA-mode WiFi APs, but also closed-access WiFi APs. From \eqref{ase} it is not obvious how the ASE improvement would evolve with the guard radius, which gives the opportunity to optimize the guard radius under specific network conditions.
From Fig. \ref{ASEuser}, we see that the ASE improvement becomes very limited when the percentage of closed-access WiFi $p$ is higher. With smaller $p$, the ASE improvement increases rapidly with the user density, meaning that the advantage of LWPA is more obvious in the dense user regime with less closed-access WiFi APs.

\begin{figure}[!t]
	\centering
	\includegraphics[width=0.9\columnwidth]{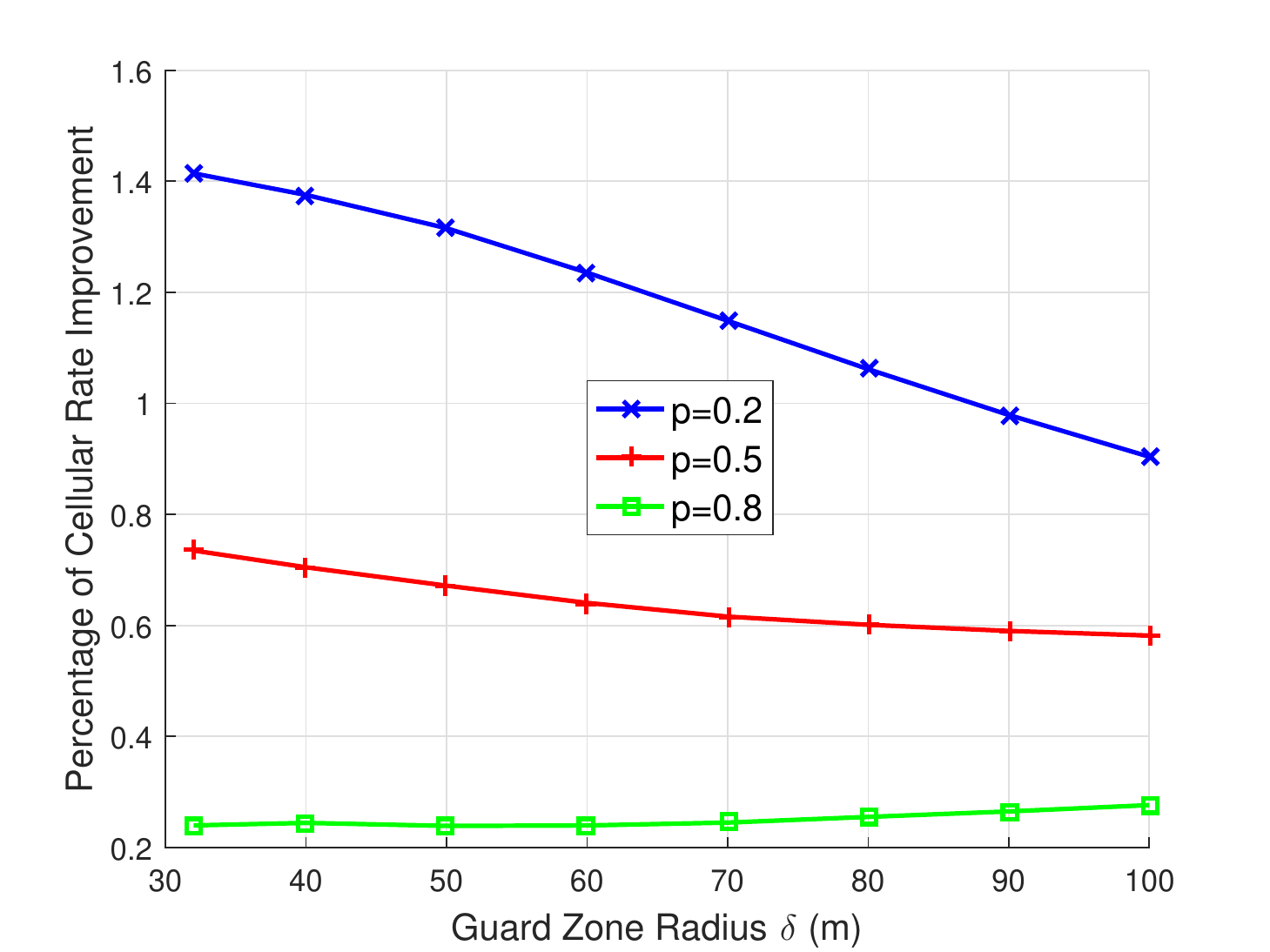}
	\caption{Cellular rate improvement vs. guard radius with $p =\{0.2, 0.5, 0.8\}$.}\label{cellularradius}
\end{figure}

\begin{figure}[!t]
	\centering
	\includegraphics[width=0.9\columnwidth]{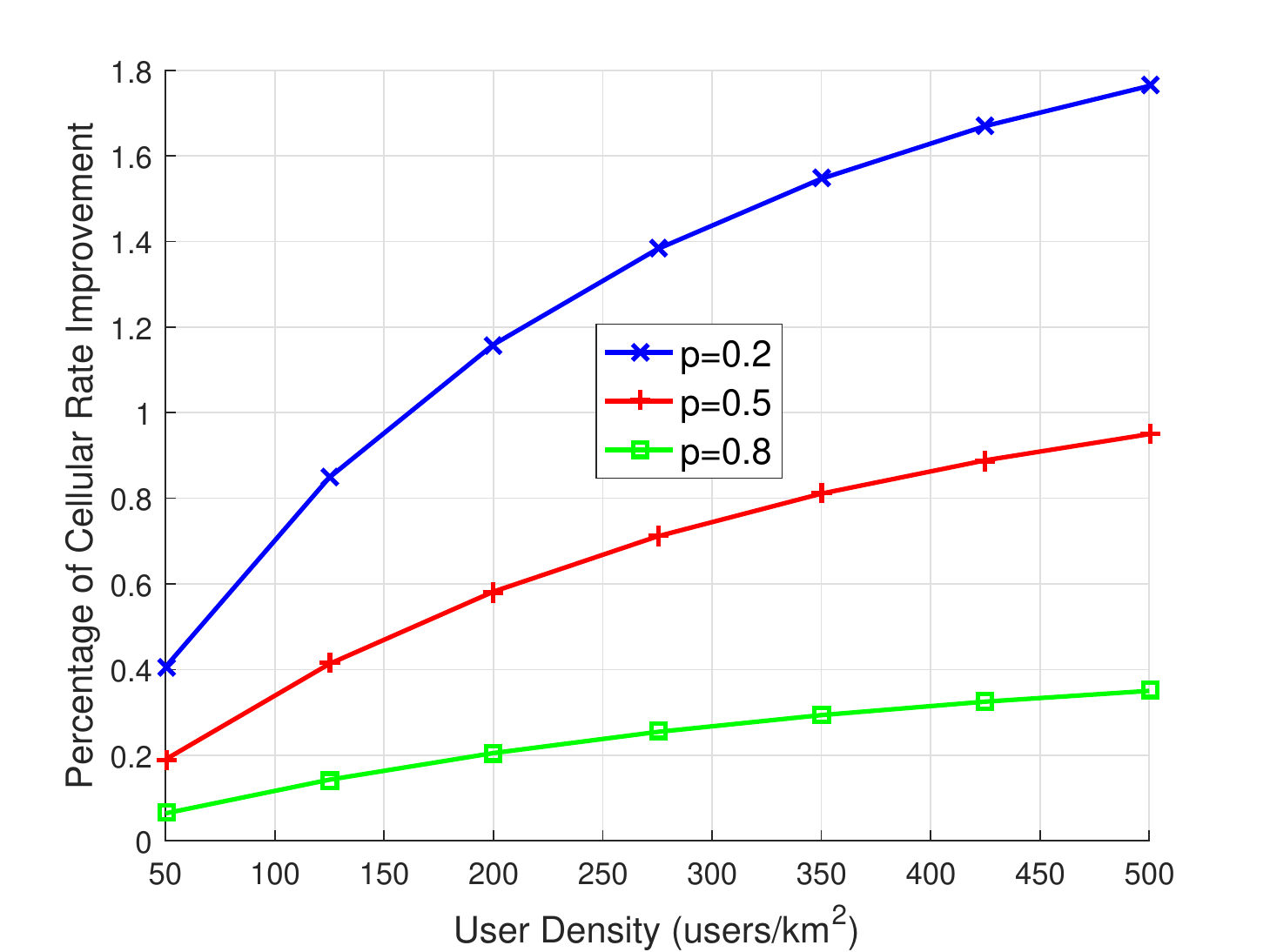}
	\caption{Cellular rate improvement vs. user density with $p =\{0.2, 0.5, 0.8\}$, and here $\delta= 50$ m.}\label{cellularuser}
\end{figure}

\subsection{Cellular Rate Improvement}
Fig. \ref{cellularradius} and Fig. \ref{cellularuser} present the percentage of cellular rate improvement versus the guard radius $\delta$ and versus the user density, respectively. The results are generated from (\ref{cellularrate}). Similar to the trend observed for WiFi ASE improvement, with smaller value of $p$, the data rate received by the cellular users can be further improved. It is worth noticing that enabling LWPA can significantly improve the downloading rates of the LTE users without causing much interference to the existing closed-access WiFi users.

\section{Conclusion}
In this paper, a stochastic geometry model was used to model and analyze the performance of an MPTCP Proxy-based LWPA network with intra-tier and cross-tier dependences. Under the considered network model and the activation conditions of LWPA-mode WiFi, we obtained three approximations for the density of active LWPA-mode WiFi APs through different approaches. Performance metrics including the success probability, the cellular rate improvement and the area spectral efficiency have been analytically derived and numerically evaluated. The impact of different parameters on the network performance have been analyzed, validating the significant gain of using LWPA in terms of boosted data rate and improved spectrum reuse.

\section*{Acknowledgment}
This work was supported in part by the Swedish Foundation for Strategic Research (SSF), the Swedish Research Council (VR), ELLIIT, the joint research project DECADE (Deploying High Capacity Dense Small Cell Heterogeneous Networks), within the Research and Innovation Staff Exchange (RISE) scheme of the European Horizon 2020 Framework Program, under contract number 645705. In addition, this work has been partially supported by the European Union's Horizon 2020 research and innovation programme under the Marie Sklodowska-Curie grant agreement No. 643002 (ACT5G).

\bibliographystyle{IEEEtran}
\bibliography{reference}
\end{document}